\documentclass[conference]{IEEEtran}
\usepackage{helvet}       
\usepackage{courier}     
\usepackage{makeidx}        
\usepackage{graphicx}                                            
\usepackage{multicol}         
\usepackage{caption}
\usepackage{amssymb, amsmath,dsfont}

\usepackage{comment}

\usepackage{cite}
\usepackage{color}
\usepackage{multirow}
\usepackage{subcaption}
\usepackage{array}
\usepackage{algorithmic}
\usepackage{algorithm}
\makeindex                                

\begin{document}
\title{Optimal Resource Allocation in Ultra-low Power Fog-computing SWIPT-based Networks}

\author{\IEEEauthorblockN{Nafiseh Janatian, Ivan Stupia and Luc Vandendorpe}
	\IEEEauthorblockA{ Universit\'e catholique de Louvain,
Louvain-la-Neuve, Belgium\\Email: \{nafiseh.janatian, ivan.stupia, luc.vandendorpe\}@uclouvain.be}}


%Telephone: (800) 555--1212, Fax: (888) 555--1212}


\maketitle

\begin{abstract}
In this paper, we consider a fog computing system consisting of a multi-antenna access point (AP), an ultra-low power (ULP) single antenna device and a fog server. The ULP device is assumed to be capable of both energy harvesting (EH) and information decoding (ID) using a time-switching simultaneous wireless information and power transfer (SWIPT) scheme. The ULP device deploys the 
harvested energy  for ID and either local computing or offloading the computations to the fog server depending on which strategy is  most energy efficient. 
In this scenario, we optimize the time slots devoted to EH, ID and local computation as well as the  time slot and power required for the offloading to minimize the energy cost of the ULP device. Numerical results are provided to study the effectiveness of the optimized fog computing system and the relevant challenges. 
\end{abstract}


% \fi
\IEEEpeerreviewmaketitle

\section{Introduction}
With the development of early 5G systems, the wireless industry finds itself at a turning point. As a matter of fact, while smartphone market penetration is saturating, new services and applications will be offered to connect human beings and \emph{things}. To achieve this, a technological breakthrough towards a new generation of low-energy mobile devices with enhanced processing capability is required. Interestingly, most of the solutions proposed so far seem to indicate an upcoming paradigm shift from traditional wireless technologies focusing on the communication aspects only, to a new framework based on a combination of \emph{computation} and \emph{communication} \cite{Mammela2017,Barb2014}. In recent years, computation offloading through cloud computing, i. e., a model for enabling ubiquitous {and} on-demand network access to a shared pool of configurable computing resources \cite{cloud}, has attracted a lot of attention.  However, as the number of connected devices increases, the way forward will be to decentralize {the computation facilities} away from the cloud towards the edge of the network closer to the user. This reduces the latency of communication between a user device and the cloud, and is the premise of \emph{fog computing} \cite{fog}.

Concurrently, new fields of investigation are stimulated by the emergence of (quasi-)autonomous ultra-low power (ULP) wireless nodes \cite{Lu15,Mi15} extending the functionality of traditional active RFID systems. The main idea behind this future generation of ULP devices is the ability of taking advantage of the same electromagnetic field for simultaneously receiving information and harvesting energy thanks to simultaneous wireless information and power transfer (SWIPT) techniques \cite{bi}.  The SWIPT concept has recently attracted a significant attention with a still growing scientific literature indicating it as an essential part for many commercial and industrial wireless systems in the future, including the IoT, wireless sensor networks and small-cell networks \cite{bi}. 
%A considerable effort has been devoted to investigate different practical SWIPT receiver architectures. 
Currently, the two practical methods
to co-locate information decoder and energy harvester in SWIPT systems are time-switching (TS)
and power-splitting (PS). In a TS design, each reception time
frame is divided into two orthogonal time slots: one for information decoding (ID) and the other for energy harvesting (EH). However, in PS design the
receiver splits the received  signal into two streams of different power levels for EH and ID \cite{zhang}.

Future ULP nodes may then embed an RF energy harvesting system to collect the energy required to perform mobile computing and wireless communication. Despite this growing interest, there are concerns about {the signal strength of far-field RF transmission that might be greatly impaired by the path-loss} when the separation between the transmitter and the RF energy harvester increases. Hence, in designing SWIPT-based ULP devices, a basic question is whether computation can be performed locally through a microcontroller with limited processing capabilities or whether computation offloading to nearby {fog} servers is more desirable. %To the best of our knowledge this investigation is still missing in the literature.  

{The integration of wireless power transfer and computation offloading technologies was recently studied in \cite{You2016, Mao2016 }. In \cite{You2016}, the authors considered an energy efficient mobile cloud computing system powered by wireless energy transfer, comprising one single-antenna mobile and one multi-antenna BS. The mobile has two operating modes, i. e., offloading and local computation, and it is assumed that the local computing and energy harvesting can be performed simultaneously.
In \cite{Mao2016} a mobile edge computing (MEC) system consisting of an EH mobile device and an MEC server is considered. Offloading strategy is optimized using the execution cost metric which addresses both the execution delay and task failure.  It is also assumed that at each time slot, a portion of the total received energy is harvested and stored in a battery.}

In this paper, different from previous works, we consider a fog computing system including a multi-antenna access point (AP), ULP single antenna device and a fog server. The ULP device is assumed to have TS SWIPT architecture which can be implemented using simple switches. ULP device deploys the harvested energy for ID and either to perform local computing or to offload the computations to the fog server.  Therefore, each time frame is divided into three orthogonal time slots, one for EH, one for ID and the other for local computation or offloading. In this scenario, we optimize the time slot and offloading power allocation to minimize the energy cost of the ULP device. The effectiveness and challenges of this framework are studied by means of numerical results.

%% no \IEEEPARstart
%% You must have at least 2 lines in the paragraph with the drop letter
%% (should never be an issue)


The rest of this paper is organized as follows. Section II describes the system model. Resource allocation and optimal offloading strategy are addressed in section III. In Section IV, we present numerical results and finally the paper is concluded in Section V.

\section{System Model}
\begin{figure}[!t]
	\centering
	\includegraphics[width=3in]{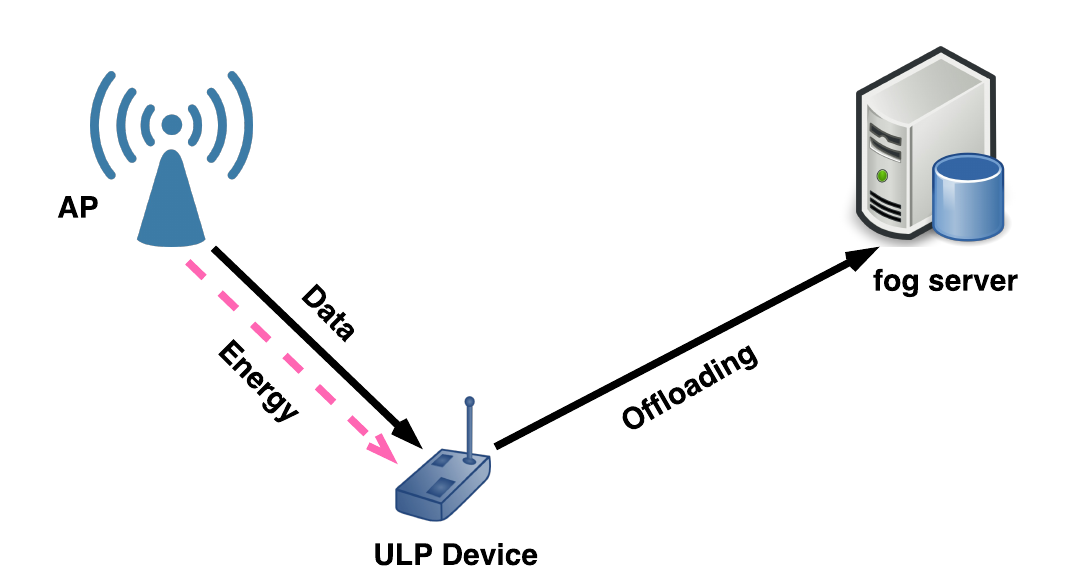}
	\caption{Fog computing system scheme}
	\label{fig1}
\end{figure}

\begin{figure}[!t]
	\centering
	\includegraphics[width=2.5in]{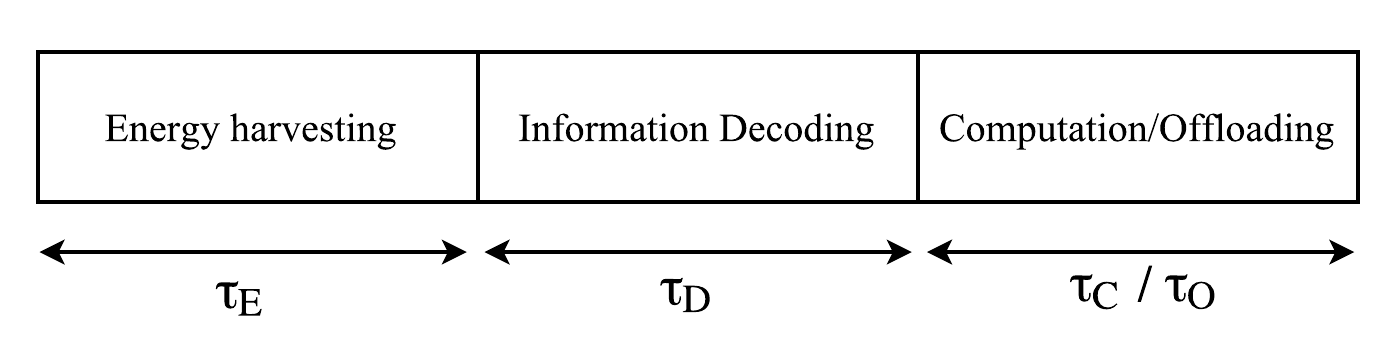}
	\caption{Time frame structure}
	\label{fig2}
\end{figure}

We consider a SWIPT fog computing system model, as presented  in Fig. \ref{fig1}, which consists of one single antenna ULP device, one AP which is equipped with $N_A$ antennas and a fog server. The ULP device is assumed to be capable of information decoding and also energy harvesting using a TS SWIPT scheme. Moreover, it decides to carry out the computation locally or to offload them to the fog server depending on the energy cost of the strategy.  As shown in Fig. \ref{fig2}, we assume that each time frame, $T$,  is divided into three orthogonal time slots: one for EH ($\tau_E$), one for ID ($\tau_D$) and the other for local computation ($\tau_C$) or offloading ($\tau_O$).  
The received signal in the ULP device can be modelled as:
\begin{equation} \label{M}
y=\boldsymbol{h^H}\boldsymbol{w}s+ n,
\end{equation}
where $s$ is the information symbol from the AP to the ULP device which originates from independent Gaussian codebooks, ${s}\sim{\mathcal{CN}(\boldsymbol{0},1)}$ and  $\boldsymbol{w}\in\mathbb{C}^{N_{A}\times1}$ is the beamforming vector. We assume quasi-static flat fading channel and denote by  $\boldsymbol{h}\in\mathbb{C}^{N_{A}\times1}$ the complex channel vector from the AP to the ULP device. Also $n\sim{\mathcal{CN}(0,\sigma_n^2)}$ is the circularly symmetric complex Gaussian receiver noise.
According to \eqref{M} and assuming perfect channel state information (CSI),  the achievable throughput $R$ (bits/sec) for the ULP device can be found as follows:
\begin{equation} \label{R}
\begin{aligned}
R=B_h\frac{\tau_D}{T}\log_2(1+\frac{|\boldsymbol{h}^Hw|^2}{\sigma_n^2}),
\end{aligned}
\end{equation}
where $B_h$ is the bandwidth of the channel between the AP and the ULP device and $\tau_D$ is the decoding time slot duration. 
In addition,  the harvested energy can be given as:
\begin{equation} \label{E}
\begin{aligned}
E_H=\eta(|\boldsymbol{h}^Hw|^2+\sigma_n^2)\tau_E,
\end{aligned}
\end{equation}
where $\eta$ denotes the energy harvesting efficiency factor for the ULP device.

As mentioned, the ULP device offloads the computations to the fog server if it is less costly than the local computation strategy. In this work, cost is defined by taking into account the amount of  harvested energy as well as the consumed energy which itself includes the two terms, i. e., decoding energy consumption $E_D$ and offloading or computation energy consumption $E_C$. \\
The amount of decoding energy consumption per bit depends on the decoding method, e. g., analogue or digital techniques and the CMOS circuit design \cite{Meraji}. Therefore, the decoding energy consumption can be written  as follows:
\begin{equation} \label{Ed}
\begin{aligned}
E_D={\epsilon}B_h\log_2(1+\frac{|\boldsymbol{h}^Hw|^2}{\sigma_n^2})\tau_D,
\end{aligned}
\end{equation}
in which $\epsilon$ (Joule/bit) is a constant  depending on the technology and circuit design. 

 On the other side, the minimum dynamic switching energy per logic gate is ${C_g}V_{DD}^2$ where $C_g$ is the gate input capacitance and $V_{DD}$ is the supply voltage \cite{Rabaey}. 
This energy consumption is estimated to be above the Landauer limit by a factor of $M_c$, i. e., $M_cN_0\ln(2)$ \cite{Mammela2017}, where $M_c$ is a time-dependent immaturity factor of the technology and $N_0$ is the thermal noise spectral density.  Deploying this estimation, the computation energy consumption can be modelled as:
\begin{equation} \label{Ec}
\begin{aligned}
E_C=F_0{\alpha}M_cN_0\ln(2)KRT,
\end{aligned}
\end{equation}
where $F_0$ is the fanout, which is the number of loading logic gates, typically 3-4, $\alpha$ is the activity factor which is typically 0.1-0.2, $K$ is the number of logic operations per bit and $RT$ is the number of received bits during each time frame.\\
In case of offloading, the received bits are sent to the fog server across the channel link from the ULP device to the fog server with the bandwidth of $B_g$. Therefore, the number of bits that can be offloaded during the time $\tau_O$ is given by:
\begin{equation} \label{No}
\begin{aligned}
N_O=B_g{\tau_O}\log_2(1+\frac{|g|^2p_O}{\sigma_s^2}),
\end{aligned}
\end{equation}
where $g$ is the complex-valued channel coefficient from the ULP device to the fog server, $p_O$ is the power spent for transmitting data to the server and $\sigma_s^2$ is the receiver's noise power. 
Thereby, we can define the system's cost function as follows:
\begin{equation} \label{C}
\begin{aligned}
&C(\boldsymbol{\tau},p_O, I_O)=
&E_C+({\tau_O}p_O-E_C)I_O+E_D-E_H,
\end{aligned}
\end{equation}
in which $\boldsymbol{\tau}=[\tau_E,\tau_D,\tau_C,\tau_O]$ and $I_O\in\{0,1\}$ is the offloading indicator, i.e., we offload the data if $I_O=1$, otherwise the computation will be done locally.
%\begin{equation} \label{Io}
%\begin{aligned}
%I_{O}=u(C|_{I_O=0}-C|_{I_O=1}),
%\end{aligned}
%\end{equation}
%where $u(.)$ is the unit step function.
Negative values of $C(.)$  are related to the cases in which decoding, computation or offloading processes could be done and the additional part of the harvested energy could be stored. However, positive values of $C(.)$ in each frame indicate that the energy consumption is more than the harvested energy in that frame. 
As a consequence, if the stored energy at the beginning of this frame can not compensate the cost, the process including decoding, computation or offloading can not be conducted. In this case the ULP device only harvests energy, i. e., $\tau_E=T$. Consequently, the  energy storage at the beginning of $(i+1)$th time frame can be written as follows:
 \begin{equation} \label{Io}
 \begin{aligned}
E_s^{(i+1)}=E_s^{(i)}-(1-I_s^{(i)})C^{(i)}+I_{s}^{(i)}E_H^{(i)}|_{\tau_E=T},
 \end{aligned}
 \end{equation}
in which $C^{(i)}$ and $E_H^{(i)}$ are the cost and harvested energy in the $i$th time frame, respectively.  $I_s^{(i)}$ indicates whether the ULP device is in harvesting mode  or can also carry out the decoding, computation or offloading that is:
 \begin{equation}
I_s^{(i)}=
\begin{cases}
0, & \text{if}\  E_s^{(i)}-C^{(i)} \geq 0 \\
1, & \text{otherwise}
\end{cases}
\end{equation}

\section{Optimal resource allocation and offloading decision}

In this section, we optimize the time slot  and offloading power allocation to minimize the energy cost at the side of the ULP device. Moreover, we make the final decision whether to carry out the computations locally or to offload the data to the fog server by comparing the optimal costs  of the two cases. The optimization problem for the $i$th time frame is formulated as follows:

\begin{equation} \label{P1} 
\begin{aligned}
&\underset{\boldsymbol{\tau},p_O,I_O}{\text{Minimize}}
& &C(\boldsymbol{\tau},p_O, I_O)\\
& \text{subject to} 
& & \mbox{(1)}  \ R\geq\bar{R}\\
&
& & \mbox{(2)}  \ KN_OI_O+\tau_Cf_{op}(1-I_O)\geq {KRT} \\
&
& & \mbox{(3)}  \   C\leq{E_s}\\
&
& & \mbox{(4)}  \    \tau_E+\tau_D+\tau_C+I_O(\tau_O-\tau_C)=T\\
&
& & \mbox{(5)} \boldsymbol{\tau}\in[0,T],   \ I_{O}\in\{0,1\},
\end{aligned}
\end{equation}
in which  the $i$ superscript for variables are omitted for brevity, i. e., $C(.)$ is the cost function for the $i$th frame and $E_s$ is the given stored energy at the beginning of this frame.
The first constraint in the above problem is the minimum throughput requirement of the ULP device. In the second constraint,  if $I_O=1$ (the best strategy is to offload the data) then $N_O\geq{RT}$,  implying that the number of offloaded bits must be greater than or equal to the number of received bits in one frame. However, if $I_O=0$ then $KRT\leq{f_{op}\tau_C}$ ($f_{op}$ is  the maximum number of the operations per second  in the ULP device), which means that the total number of logic operations must be less than the computational capacity of the ULP device during computation time. The third constraint guarantees that we have enough energy for the whole process. In the following we relax this constraint and find the optimal solution of the relaxed problem. Obviously, if $C^*>E_s$ , the ULP goes to harvesting mode ($\tau_E^*=T$) as mentioned before. 
\subsection{Optimal resource allocation for local computation}
In the case of local computation, problem \eqref{P1} will be simplified to the following problem:
\begin{equation} \label{P2} 
\begin{aligned}
&\underset{\tau_E,\tau_D,\tau_C}{\text{Minimize}}
& &E_D+E_C-E_H\\
& \text{subject to} 
& & \mbox{(1)}  \ R\geq\bar{R},\\
&
& & \mbox{(2)}  \ \tau_Cf_{op}\geq {KRT}, \\
&
& & \mbox{(3)}  \    \tau_E+\tau_D+\tau_C=T,\\
&
& & \mbox{(4)}  \ {\tau_C, \tau_D, \tau_E}\in[0,T],
\end{aligned}
\end{equation}
which after substitution of $E_C$, $E_D$, $E_H$ and $R$ yields to:
 \begin{equation} \label{P3} 
 \begin{aligned}
 &\underset{\tau_D,\tau_C}{\text{Minimize}}
 & &\tau_DB_h\log_2(1+\frac{|\boldsymbol{h}^Hw|^2}{\sigma_n^2})(KF_0{\alpha}M_cN_0\ln2+\epsilon)-\\
 &
 &&\eta(|\boldsymbol{h}^Hw|^2+\sigma_n^2)(T-\tau_D-\tau_C),\\
 & \text{subject to} 
 & & \mbox{(1)}  \ B_h\frac{\tau_D}{T}\log_2(1+\frac{|\boldsymbol{h}^Hw|^2}{\sigma_n^2})\geq\bar{R},\\
 &
 & & \mbox{(2)}  \ \tau_Cf_{op}\geq {K\tau_DB_h\log_2(1+\frac{|\boldsymbol{h}^Hw|^2}{\sigma_n^2})}\\
 &
 & & \mbox{(3)}  \ {\tau_C, \tau_D}\in[0,T].
 \end{aligned}
 \end{equation}
 According to the first and second constraints and the fact that $\tau_C+\tau_D\leq{T}$,  the above problem is feasible only if:
 \begin{equation} \label{feas}
\frac{1}{B_h\log_2(1+\frac{|\boldsymbol{h}^Hw|^2}{\sigma_n^2})}+\frac{K}{f_{op}}
\leq{\frac{1}{\bar{R}}}.
 \end{equation}
 Here we assume that $\bar{R}$, $K$ and $f_{op}$ are chosen such that \eqref{feas} is satisfied. 
 
The objective function in problem \eqref{P3} is an increasing function of $\tau_C$ and $\tau_D$. As a result, the optimal $\tau_D$ and $\tau_C$ occur when the two constraints meet their boundaries. The optimal time allocation is therefore as below:
\begin{align} \label{TD}
\tau_D^*&=\frac{\bar{R}T}{B_h\log_2(1+\frac{|\boldsymbol{h}^Hw|^2}{\sigma_n^2})},\\
\tau_C^*&=\frac{K\bar{R}T}{f_{op}},\\
\tau_E^*&=T-\tau_D^*-\tau_C^*.
\end{align}

\subsection{Optimal resource allocation for offloading}
In the case of offloading, problem \eqref{P1} can be written as:
\begin{equation} \label{P4} 
\begin{aligned}
&\underset{\tau_E,\tau_D,\tau_O, p_O}{\text{Minimize}}
& &E_D+\tau_Op_O-E_H,\\
& \text{subject to} 
& & \mbox{(1)}  \ R\geq\bar{R},\\
&
& & \mbox{(2)}  \ N_O\geq {RT}, \\
&
& & \mbox{(3)}  \    \tau_E+\tau_D+\tau_O=T,\\
&
& & \mbox{(4)}  \ {\tau_O, \tau_D, \tau_E}\in[0,T].
\end{aligned}
\end{equation}
According to the first constraint and the fact that  $\tau_D<T$, this problem is feasible if
$\bar{R}<{B_h\log_2(1+\frac{|\boldsymbol{h}^Hw|^2}{\sigma_n^2})}$. Here we assume that $\bar{R}$ is chosen such that the problem is feasible. 
Substitution of $E_D$, $E_H$, $R$ and $N_O$ in the above problem leads to:\\
\begin{equation} \label{P5} 
\begin{aligned}
&\underset{\tau_E,\tau_D,\tau_O,p_O}{\text{Minimize}}
&&\epsilon\tau_DB_h\log_2(1+\frac{|\boldsymbol{h}^Hw|^2}{\sigma_n^2})+\tau_Op_O-\\
&
&&\eta(|\boldsymbol{h}^Hw|^2+\sigma_n^2)\tau_E,\\
& \text{subject to} 
& & \mbox{(1)}  \ B_h\frac{\tau_D}{T}\log_2(1+\frac{|\boldsymbol{h}^Hw|^2}{\sigma_n^2})\geq\bar{R},\\
&
& & \mbox{(2)}  \ B_g{\tau_O}\log_2(1+\frac{|g|^2p_O}{\sigma_s^2})\geq\\
&
&& \ \ \ \ {B_h{\tau_D}\log_2(1+\frac{|\boldsymbol{h}^Hw|^2}{\sigma_n^2})},\\
&
& & \mbox{(3)}  \    \tau_E+\tau_D+\tau_O=T,\\
&
& & \mbox{(4)}  \ {\tau_O, \tau_D, \tau_E}\in[0,T].
\end{aligned}
\end{equation}
Due to the same reason as in the previous case, the optimal decoding time in this strategy can be found from equation \eqref{TD} as well. Therefore, the problem can be further simplified to:
 \begin{equation} \label{P6} 
 \begin{aligned}
 &\underset{\tau_E,\tau_O,p_O}{\text{Minimize}}
 &&\tau_Op_O-\eta(|\boldsymbol{h}^Hw|^2+\sigma_n^2)\tau_E\\
 & \text{subject to} 
 & & \mbox{(1)}  \ B_g{\tau_O}\log_2(1+\frac{|g|^2p_O}{\sigma_s^2})\geq \bar{R},\\
 &
 & & \mbox{(2)}  \    \tau_E+\tau_D^*+\tau_O=T,\\
 &
 & & \mbox{(3)}  \ {\tau_O, \tau_D, \tau_E}\in[0,T],
 \end{aligned}
 \end{equation}
and by introducing the slack variable $\lambda=p_O\tau_O$ we have:
 \begin{equation} \label{P7} 
\begin{aligned}
&\underset{\tau_E,\tau_O,\lambda}{\text{Minimize}}
&&\lambda-\eta(|\boldsymbol{h}^Hw|^2+\sigma_n^2)\tau_E\\
& \text{subject to} 
& & \mbox{(1)}  \ B_g{\tau_O}\log_2(1+\frac{|g|^2\lambda}{\tau_O\sigma_s^2})\geq\bar{R},\\
&
& & \mbox{(2)}  \    \tau_E+\tau_D^*+\tau_O=T,\\
&
& & \mbox{(3)}  \ {\tau_O,  \tau_E}\in[0,T].
\end{aligned}
\end{equation}
The objective function and the second constraint in the above problem are linear functions of the optimization variables. Also it can be easily shown that the left side of the inequality in the first constraint is a concave function of both $\tau_O$ and $\lambda$. 
Using the KKT conditions of the above problem and after some mathematical manipulations, the optimal offloading time can be found as:
\begin{align} \label{TO}
\tau_O^*&=\frac{\bar{R}T\ln(2)}{B_g}\left(1+\mathcal{W}\left(\frac{1}{e}(\frac{\sigma_n^2}{\sigma_s^2}\eta|g|^22^{\frac{\bar{R}T}{B_h\tau_D^*}}-1)\right)\right)^{-1},
\end{align}
where $\mathcal{W}(.)$ is the Lambert function \cite{lambert}, i.e., the inverse function of $f(x)=xe^x$. The optimal offloading power and harvesting duration can then be calculated from the following equations:
\begin{align} \label{TO2}
p_O^*&=\frac{\sigma_s^2}{|g|^2}(2^{\frac{\bar{R}T}{B_g\tau_O^*}}-1),\\
\tau_E^*&=T-\tau_D^*-\tau_O^*.
\end{align}
\subsection{Final strategy}
According to the discussion above, the optimal decoding time and therefore the optimal decoding energy consumption will be the same for both cases under the same throughput requirement. Denote the optimal cost of the local computing and offloading strategy by $C_1^*$ and $C_2^*$, respectively. Therefore if $\min(C_1^*,C_2^*)\leq{E_s}$,  the decision making strategy would be:
 \begin{equation} \label{Io*2}
\begin{aligned}
I_{O}^*=&u(C_1^*-C_2^*),
%=\\
%&u((E_D^*+E_C^*-{E_H^*}_1)-(E_D^*+\tau_O^*p_O^*-{E_H^*}_2)),
\end{aligned}
\end{equation}
%where the optimal energy values of the local computing strategy and offloading strategyare denoted  by $E_D^*$,  $E_C^*$, ${E_H^*}_1$ and $E_D^*$,  $\tau_O^*p_O^*$, ${E_H^*}_2$, respectively and
 where $u(.)$ is the unit step function.  Otherwise, the ULP device will only harvest energy  ($\tau_E^*=T$).
After substituting the optimal costs in \eqref{Io*2}, the final strategy can be written as:
\begin{equation} \label{Io*3}
\begin{aligned}
K\bar{R}T(F_0{\alpha}M_cN_0\ln(2)+\frac{\eta(|\boldsymbol{h}^Hw|^2+\sigma_n^2)}{f_{op}})\overset{I_O=1}{\underset{I_O=0}\gtrless}\\\tau_O^*\left(\frac{\sigma_s^2}{|g|^2}(2^{\frac{\bar{R}T}{B_g\tau_O^*}}-1)+\eta(|\boldsymbol{h}^Hw|^2+\sigma_n^2)\right),
\end{aligned}
\end{equation}
in which $\tau_O^*$ is defined by \eqref{TO}. For given values of $\bar{R}$,  $T$, noise powers  and  computational parameters, the final result depends only on the transmit and offload channel realizations of $\boldsymbol{h}$ and $g$ and the bandwidth of the offloading channel, $B_g$, as it is expected.

\section{Numerical Results}
In this section, we present numerical results to demonstrate the performance of our fog computing SWIPT system. We consider a single user scenario with the same setup as in Fig. \ref{fig1} in an ultra dense network, consisting of one AP equipped with $N_A$ = 4 antennas and a single antenna TS SWIPT ULP device. 
The line-of-sight (LoS) component is dominant in the short distances, thus the channel gains $\boldsymbol{h}$ from the AP to the ULP device and $\boldsymbol{g}$ from the ULP device to the fog server are generated with Rician fading. The Rician factor, defined as the
ratio of signal power in dominant component over the scattered power, is set to 3.5 dB. Besides, path-loss of the channels is modelled using the ITU indoor channel model  as \cite{ITU}:
 \begin{equation} \label{Io*}
\begin{aligned}
L=20\log f_c+N\log d-28,
\end{aligned}
\end{equation}
in which $L$ (dB) is the total path loss, $f_c$ (MHz) is the frequency of transmission, $d$ (m) is the distance and $N$ is the distance power loss coefficient. We have set $f_c=2.4$ GHz and $N=22$ which is the predicted value for the commercial area \cite{ITU}.
Noise powers are assumed to be $\sigma_n^2=\sigma_s^2=-110$ dB. Conjugated beamforming is used, i. e., $\boldsymbol{w}=\sqrt{P_T}[e^{-j\Phi_1}, e^{-j\Phi_2},...,e^{-j\Phi_{N_A}}]^T$, where $\Phi$ is the $N_A\times{1}$  phase vector related to the $\boldsymbol{h}$ and $P_T$ is the AP's power budget which is set to 1 watt.
Both channels are assumed to have the bandwidth of $B_h=B_g=2$ MHz and $10^3$ realizations of the channel are used for averaging. Moreover, the minimum required throughput is set to $\bar{R}=20$ Kb/s, time frame duration of $T=1$ sec is considered and energy harvesting efficiency of the ULP device is assumed to be $\eta=0.6$. Also the decoding energy consumption per bit is chosen to be $\epsilon=100$ pJ/bit \cite{Meraji}. The parameters relating to the computation are $f_{op}=10^9$ operations per second, $M_c=10^4$, $\alpha=0.1$ and $F_0=3$ \cite{Mammela2017}. 

\begin{figure}[!t]
	\centering
	\includegraphics[width=3.2in]{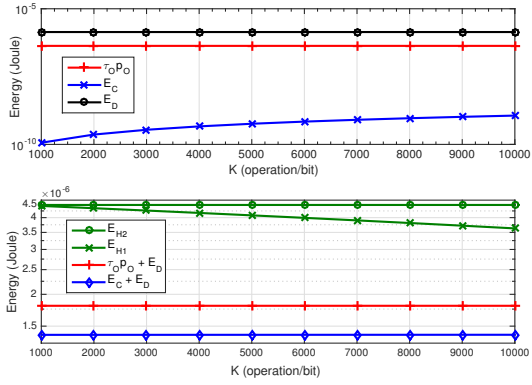}
	\caption{Consumed and harvested energies  per  frame versus $K$}
	\vspace{-.3cm}
	\label{sim3}
\end{figure}

First, we study the effect of the number of operations per bit, $K$, on the optimal energy consumptions per frame.  Fig. \ref{sim3}  shows the average consumed and harvested energies versus $K$ for a fixed distance of $d_t=6$ m from the AP to the ULP device and $d_s=10$ m from the ULP device to the fog server. As can be inferred from equation \eqref{TD} and also can be seen in Fig. \ref{sim3} (top), as far as the required throughput is fixed, $E_D$ does not change.
The offloading energy consumption, $\tau_Op_O$, does not  change with increasing $K$ as well and is fixed while $d_s$ is fixed.
Also as shown, in this scenario $E_C$ is much lower than decoding and offloading energy by few orders of magnitude. However, it increases linearly (notice the logarithmic scale of y-axis) with increasing the number of operations per bit. 

\begin{figure}[!t]
	\centering
	\includegraphics[width=3in]{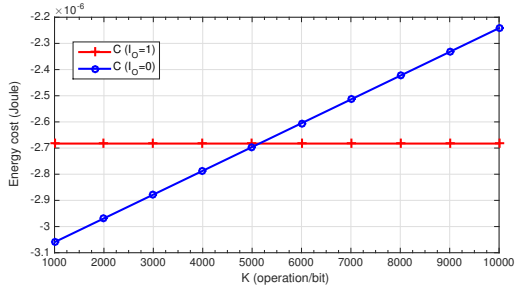}
	\caption{Energy costs per  frame versus $K$}
		\vspace{-.3cm}
	\label{sim5}
\end{figure}

\begin{figure}[!t]
	\centering
	\includegraphics[width=3in]{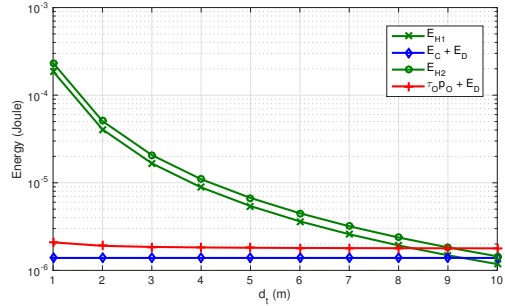}
	\caption{Energy consumption and harvested energy per  frame versus $d_t$}
	\label{sim1}
\end{figure}

The average of the total optimal energy consumptions in both local computation and offloading strategies is also illustrated in Fig. \ref{sim3} (bottom) as well as the average optimal harvested energy in these two cases, which are denoted by $E_{H1}$ and $E_{H2}$, respectively.  As can be seen, by increasing $K$, the amount of harvested energy decreases, since longer time will be devoted to the computation which regarding the fixed decoding time leads to less time for harvesting energy. Finally, the energy costs of the two strategies are shown in Fig. \ref{sim5}. According to these costs, up to nearly $K=5000$ operations per bit, we have, in average, less energy cost in local computation strategy and after that point offloading is preferred.  
%\begin{figure}[!t]
%	\centering
%	\includegraphics[width=3in]{sim4.eps}
%	\caption{Energy consumption and harvested energy per  frame versus $K$}
%		\vspace{-.3cm}
%	\label{sim4}
%\end{figure}
We have also plotted the average consumed and harvested energy  versus $d_t$ for $d_s$=10 m and $K=10^4$ operations per bit in Fig. \ref{sim1}. 
As it is shown in this figure, the optimal energy consumption including decoding and computation or offloading energies does not change with $d_t$, since  $K$ and $d_s$ are fixed. However, the amount of harvested energy decreases significantly by moving away the AP because of the channel path-loss.
 Moreover, it can be seen that  for both cases, the harvested energy covers the energy consumptions up to $d_t=9$ m  in average.  

%The demonstrated results so far,  consider the average energy costs. To illustrate the statistical performance of the system, we have defined an outage probability as the probability of the optimal energy cost being positive.  In fact, at each iteration, by solving the optimization problem \eqref{P1},  the best strategy whether the local computation or offloading is taken. Then if the energy cost  becomes negative, the additional harvested energy is saved and deployed in the next iteration, otherwise it is counted as an outage. 

\begin{figure}[!t]
	\centering
	\begin{subfigure}[b]{0.5\textwidth}
		\includegraphics[width=2.9in]{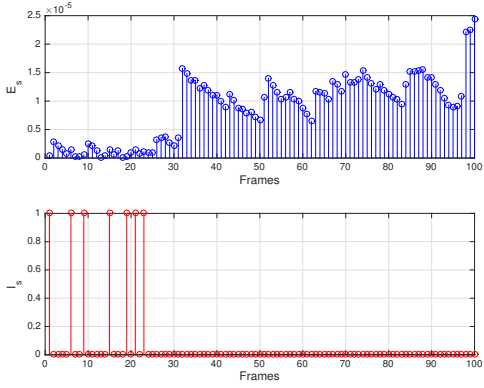}
		\caption{}
		\label{sim6a}
	\end{subfigure}\\
	\begin{subfigure}[b]{0.5\textwidth}
		\includegraphics[width=2.9in]{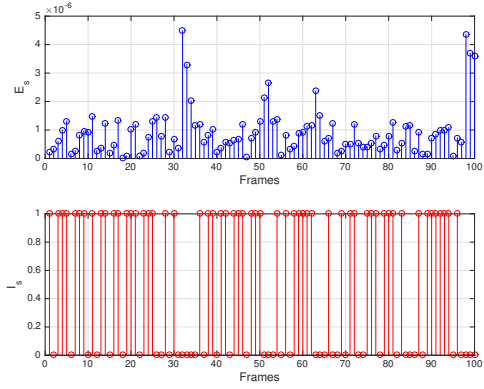}
		\caption{}
				\label{sim6b}
	\end{subfigure}
	\caption{Energy storage and $I_s$ indicator for $K=10^4$ operations/bit and (a) $d_t=10$ m and (b)  $d_t=15$ m}
		\label{sim6}
\end{figure}

\begin{figure}[!t]
%	\centering
	\includegraphics[width=3in]{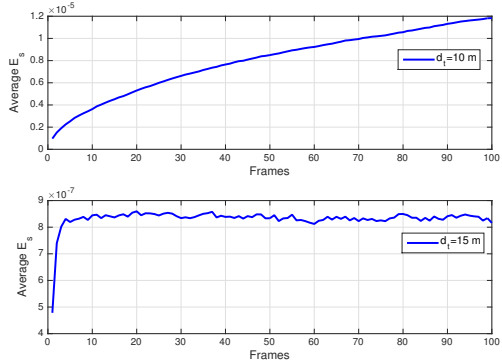}
	\caption{Average energy storage for $K=10^4$ operations/bit}
	\vspace{-.3cm}
	\label{sim7}
\end{figure}

\begin{figure}[!t]
	\centering
	\includegraphics[width=3in]{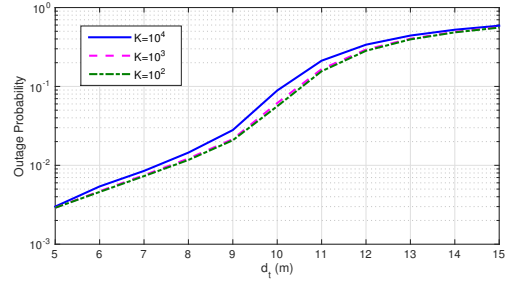}
	\caption{Outage probability versus $d_t$ for $K=10^2, 10^3, 10^4$}
		\vspace{-.3cm}
	\label{sim8}
\end{figure}

Energy storage levels and $I_s$ indicator change during 100 time frames are illustrated in Fig. \ref{sim6}  for one channel realization per each frame, $K=10^4$ operations/bit and $d_t=10, 15$ m.  As can be seen in Fig. \ref{sim6a}, since in the first frame the amount of harvested energy is not enough for completing the process including the decoding and local computing or data offloading, the ULP goes to the harvesting mode. This amount of stored energy helps the ULP to carry out the process for the next several frames, but it goes to the harvesting mode again in the 6th frame. This procedure continues until the ULP faces a strong channel in the 32nd frame and harvests enough energy so that it can decode the received signal and compute or offload the data in the remaining time frames.  However, in farther distance of $d_t=15$ m, the ULP keeps going to the harvesting mode after each 2-3 frames  due to severe path loss effect as shown in Fig. \ref{sim6b}.\\
The average energy storage levels during 100 time frames are also plotted in Fig. \ref{sim7} for $K=10^4$ operations/bit and  $d_t=10, 15$ m. As can be seen, the average stored energy increases constantly in time for $d_t=10$ m. However, in $d_t=15$ m, the energy storage level increases in the beginning and then almost saturates which was expected according to the one realization result of Fig. \ref{sim6b}.\\
Finally to illustrate the statistical performance of the system, we define the outage probability as the probability of the ULP device going to harvesting mode, i.e. the probability of $I_s=1$.  The average outage probability in 100 frames versus $d_t$  for  $K=10^2, 10^3, 10^4$ operation/bit is plotted in Fig. \ref{sim8}.  As can be seen, the probability of outage is less than 1\%  in low distances ($d_t<7$ m) , while  by increasing the distance to $d_t=15$ m, the probability of outage increases to nearly 57\%. Besides, the probability of outage is slightly higher in bigger number of operations per bit.
\section{Conclusion}
In this paper, we investigated a SWIPT-based fog computing system. We optimized the required time intervals for EH, ID and local computation  as well as the  time slot and power required for  offloading the computations to the fog server. The optimization goal has been  to minimize the energy cost  with a constraint on the minimum data rate requirement of the ULP device. We studied the effect of the number of computational operations per bit and the distance between AP and ULP device by means of numerical simulations. The results showed that, by moving the AP away from the ULP device, less energy can be harvested and as a result the probability of outage will increase. Moreover, increasing the number of computational operations can also decrease the amount of harvested energy by limiting the EH time interval.

\section*{Acknowledgment}
The authors would like to thank IAP BESTCOM project funded by BELSPO, and the FNRS for the financial support.



%\bibliographystyle{IEEEtran}

\end{document}